# Quantum oscillations between excitonic and quantum spin Hall insulators in moiré WSe$_2$

Zhongdong Han[1*†], Yiyu Xia[2*], Kenji Watanabe[3], Takashi Taniguchi[3], Kin Fai Mak[1,2,4,5†], Jie Shan[1,2,4,5†]

[1]Laboratory of Atomic and Solid State Physics, Cornell University, Ithaca, NY, USA
[2]School of Applied and Engineering Physics, Cornell University, Ithaca, NY, USA
[3]National Institute for Materials Science, Tsukuba, Japan
[4]Kavli Institute at Cornell for Nanoscale Science, Ithaca, NY, USA
[5]Max Planck Institute for the Structure and Dynamics of Matter, Hamburg, Germany

*These authors contributed equally
†Email: zh352@cornell.edu; kin-fai.mak@mpsd.mpg.de; jie.shan@mpsd.mpg.de

**Quantum spin Hall insulators (QSHIs) and excitonic insulators (EIs) are prototypical topological and correlated states of matter, respectively. The topological phase transition between the two has attracted much theoretical interest but experimental studies have been hindered by the availability of tunable materials that can access such a transition. Here, by utilizing the interaction-enhanced g-factor and the flat moiré bands in twisted bilayer WSe$_2$ (tWSe$_2$), we realize tunable electron-like and hole-like Landau levels (LLs) in the opposite valleys of tWSe$_2$ under a perpendicular magnetic field. At half-band-filling, which corresponds to electron-hole charge neutrality, periodic oscillations between QSHIs (for fully filled LLs) and EIs (for half-filled LLs) are observed due to the interplay between the cyclotron energy and the intervalley correlation; QSHIs with up to four pairs of helical edge states can be resolved. We further analyze the effect of Fermi surface nesting on the stability of EIs via electric field-tuning of the moiré band structure. Our results demonstrate a novel QSHI-to-EI topological phase transition and provide a comprehensive understanding of the fermiology of tWSe$_2$.**

## Main

Quantum spin Hall insulators (QSHIs)[1-5] are symmetry-protected topological states with an insulating bulk surrounded by pairs of helical edge states in two dimensions. Excitonic insulators (EIs)[6-12] are correlated insulating states arising from the spontaneous formation of a fluid of bound electron-hole pairs (excitons) when the band gap of an insulator becomes smaller than its exciton binding energy. The topological phase transition between these distinct insulating states provides a playground where strong correlation effects and nontrivial band topology intersect[13-16]; the transition can be realized in small band gap insulators near a band inversion critical point in the presence of strong electron-hole interactions[17-19]. Although experimental studies on InAs/GaSb quantum wells[20-25] and monolayer T$_d$-WTe$_2$ (Ref.[26-29]) have shown evidence of coexisting EIs and QSHIs, unambiguous demonstration of the topological phase transition between the two states has remained elusive because of the difficulty in controlling the band structure continuously near the inversion critical point in these materials.

In this work, we present a different approach to realize the QSHI-to-EI topological phase transition by engineering the electron-like and hole-like LLs in twisted bilayer WSe$_2$ (tWSe$_2$)[30-33]. Figures 1a and 1b show the first moiré valence band of 3.65°-twisted tWSe$_2$ in its moiré Brillouin zone

(mBZ)[34,35]. (See Methods for band structure calculations.) In the absence of an interlayer potential difference, the K- and K'-valley states, which are locked to the spins due to the strong Ising spin-orbit coupling (SOC)[36] in $WSe_2$, are degenerate. Hole doping the material first creates two hole Fermi pockets centered at $\kappa_+$ and $\kappa_-$ of the mBZ; an increase in doping density beyond the van Hove singularity (vHS) at the $m$-point of the mBZ turns the hole pockets into an electron pocket centered at $\gamma$ of the mBZ. The valley/spin degeneracy is lifted under a finite interlayer potential difference, which creates an emergent Ising SOC in the moiré band structure[37-40]. The K and K' hole pockets are now centered at $\kappa_+$ and $\kappa_-$, respectively; the electron pockets at $\gamma$ also acquire a trigonal warping correction.

The narrow bandwidth and the large valley/spin g-factor of $tWSe_2$ (the latter is further enhanced by interaction effects as we will show below) allow a full Zeeman splitting of the K- and K'-valley states by a laboratory-scale magnetic field ($B$). At half-band-filling (or moiré lattice filling factor $\nu = 1$), the application of a sufficiently high B-field would polarize all the holes to one valley (Fig. 1c); this state corresponds to a vacuum state in our setup. Lowering the B-field reduces the valley Zeeman splitting and populates the electron-LLs at $\gamma$ of the K-valley (spin up) and, simultaneously, the hole-LLs at $\kappa_-$ of the K'-valley (spin down) while keeping the system charge neutral. (K and K' would be flipped under opposite B-fields.) The B-field tunes the pair density and the pair chemical potential (which is proportional to the valley Zeeman energy). When both the electron- and hole-LLs are fully filled, QSHIs with spin-polarized helical edge states are expected[41,42] (Fig. 1d); the QSHIs here are protected by spin-$S_z$ conservation due to the Ising SOC and are characterized by the Z topological invariant[43,44] (in contrast to the $Z_2$ topological insulator protected by time-reversal symmetry[45]); therefore, multiple pairs of helical edge states are allowed[46,47] as we will demonstrate below. On the other hand, EIs may emerge when the LLs are half-filled if the electron-hole interaction is sufficiently strong[48-50] (Fig. 1d). Oscillatory transitions between EIs and QSHIs as a function of the B-field are therefore possible[48,49,51]. Note that the physics here is closely connected to the reported quantum oscillations (QOs) between EIs and layer-decoupled quantum Hall states in Coulomb-coupled $MoSe_2/WSe_2$ electron-hole double layers[52,53]; the layer degree of freedom in $MoSe_2/WSe_2$ is now replaced by the valley/spin degree of freedom in $tWSe_2$.

**Phase diagrams of $tWSe_2$ under B-field**
We fabricate dual-gated Hall bar devices of $tWSe_2$ to access its electrical transport properties under tunable filling factor for holes ($\nu$) and perpendicular electric field ($E$); the latter is proportional to the interlayer potential difference. (See Methods for device fabrication and electrical measurements.) We focus on devices with twist angles near 3.65° because too large an angle would require too high a B-field to reach full valley splitting of the moiré bands[50] and too small an angle would suffer from stronger twist angle disorder effects[54]. A specific twist angle is, however, not required for the reported phenomena below. Unless otherwise specified, the sample (lattice) temperature is at 40mK.

Figure 2a shows the longitudinal resistance ($R_{xx}$) of a 3.65° device as a function of $\nu$ and $E$ at $B = 0T$. We compare the results with the calculated electronic density of states (DOS) in Fig. 2b. As demonstrated in an earlier study[32], there is an overall good agreement between theory and experiment except the emergence of correlated insulators at $\nu = 1$, 1/3 and 1/4 and of superconductivity near $\nu = 1$ and $E = 0mV/nm$, phenomena of which cannot be captured by the band theory. In Fig. 2a, we can identify a layer-hybridized region bound by the dashed lines and layer-polarized regions above and below the dashed lines; the vHS with large DOS in the layer-hybridized region is manifested by an enhanced $R_{xx}$ traced by the red curve. We also show the Fermi surfaces (for the K-valley states only) at representative regions of the phase diagram in Fig. 2b; across the vHS, the hole pockets centered at $\kappa_+$ and $\kappa_-$ are turned into a single electron pocket at $\gamma$.

Next, we examine the behavior under $B = 8T$ (Fig. 2c). (Results at other B-fields are shown in Extended Data Fig. 1.) A schematic phase diagram summarizing the results is shown in Fig. 2d. The Fermi surface topology corresponding to the different regions of the phase diagram are shown in Fig. 2e. There are two distinct regions in the layer-polarized regime due to valley Zeeman splitting: Region I with valley- (and spin-) polarized LLs and Region II with LLs from both valleys occupied. The behaviors here are very similar to those in monolayer WSe$_2$ (Ref. [55,56]) or in the layer-polarized regime of natural bilayer WSe$_2$ (Ref. [57]). In the layer-hybridized regime, the vHS from each valley (red curves in Fig. 2d) is Zeeman split by the external B-field, resulting in four distinct regions: Regions III and IV are valley-polarized and Regions V and VI with both valleys occupied. Region III has small hole Fermi pockets; the region is dominated by the Wigner-Mott insulators[58] at $\nu = 1/3$ and $1/4$ due to the strong electronic correlations at low hole doping densities. Across the vHS of the K-valley in Region IV, only a single electron Fermi pocket remains, giving rise to non-degenerate LLs independent of the E-field, i.e. the vertical $R_{xx}$ stripes in this region in Fig. 2c. The electron pocket persists in Region V while hole pockets from the K'-valley emerge; this is the only region hosting electron- and hole-LLs simultaneously. Across the vHS of the K'-valley in Region VI, only electron pockets remain; the LLs are again nearly independent of the E-field.

We also study the continuous evolution of the fermiology with B-field in Fig. 3. The experimental $R_{xx}$ as a function of $\nu$ and $B$ under a small interlayer potential difference (at $E = 10$mV/nm) is shown in Fig. 3a; a schematic phase diagram summarizing the results is shown in Fig. 3b; the different regions in the phase diagram are numbered according to their Fermi surfaces in Fig. 2e. We emphasize two observations. First, the vHS is located near $\nu = 0.8$ under zero B-field. The large DOS at the vHS has been shown to stabilize Stoner ferromagnetism[59] in small twist angle tWSe$_2$. Although Stoner ferromagnetism disappears at larger twist angles (e.g. 3.65°), the magnetic susceptibility near the vHS remains substantially enhanced (see Extended Data Fig. 3), resulting in a large Zeeman splitting of the vHS (red curves) at small B-fields; the splitting slows down at $B \approx 1 - 2T$ and is largely saturated beyond $B \approx 6T$. Second, electron- and hole-LLs are simultaneously populated near $\nu = 1$ in Region V, realizing the situation introduced in Fig. 1c. A robust insulating state emerges at $B > 11T$, where all the LLs are fully valley-polarized by the Zeeman field; the state corresponds to the vacuum state in Fig. 1c. As B-field is lowered and the electron- and hole-LLs are populated, intriguing QOs in the $\nu = 1$ insulating state appear. (See Extended Data Fig. 2 for similar oscillations at other E-fields.)

**Oscillatory transitions between QSHIs and EIs**
To further investigate these unusual QOs, we study $R_{xx}$ as a function of $E$ and $B$ at fixed $\nu = 1$ in Fig. 3c; a schematic phase diagram is shown in Fig. 3d. We will focus on Region V from now on. In general, the $\nu = 1$ correlated insulator weakens with increasing B-field. This is consistent with mean-field calculations for a ground state with a 120-degree Néel order[37,38,40]; the correlated insulator gap has been predicted to decrease monotonically as the localized spins are canted from the in-plane to the out-of-plane direction by the B-field. When the correlated insulator is sufficiently weakened, QOs start to emerge at $B \approx 3$T and become more visible with increasing B-field. The strong insulating state at $B \gtrsim 10$T and $|E| \lesssim 30$mV/nm, which has a triangular shape in the phase diagram and survives to much higher B-fields (Extended Data Fig. 4), is a fully valley-polarized insulator (i.e. the vacuum state). (The triangular shape is a result of the increased moiré bandwidth with E-field so that a larger Zeeman splitting is required to achieve full valley polarization.) As B-field decreases from the vacuum state, the system enters sequentially into regions with $\nu_{LL}^{(h)} = -\nu_{LL}^{(e)} = 1,2,3 ...$, where $R_{xx}$ plateaus near the quantized values $\frac{h}{2\nu_{LL}^{(h)} e^2}$ (see below). (Here $\nu_{LL}^{(h)}$ and $\nu_{LL}^{(e)}$ denote the hole- and electron-LL filling factor, respectively; $h$ and $e$ denote the Planck's constant and the electron charge, respectively.) These integer-filled regions also show nearly vanishing Hall resistance $R_{xy}$ (Extended Data Fig. 5). Interestingly, the integer-filled regions are separated by insulating regions (with diverging $R_{xx}$) at the partially filled LLs; the insulating states become the most robust at half-filling of the electron- and hole-LLs.

We characterize the integer-filled and the partially filled states in Fig. 4 and 5, respectively. Figure 4a shows a representative B-field line cut from Fig. 3c (at $E = 5$mV/nm) for both $R_{xx}$ and $R_{xy}$. Nearly quantized $R_{xx} \approx \frac{h}{2\nu_{LL}^{(h)} e^2}$ plateaus accompanied by small $R_{xy}$ are observed at $\nu_{LL}^{(h)} = -\nu_{LL}^{(e)} = 1,2,3$. (A weak $\nu_{LL}^{(h)} = -\nu_{LL}^{(e)} = 4$ state is also discernable.) In addition to the plateaus in B-field, nearly quantized plateaus in E-field are also observed at $\nu_{LL}^{(h)} = -\nu_{LL}^{(e)} = 1,2$ in Fig. 4b. The quantization is well developed below about 2K (Fig. 4c). These quantized states are separated by regions with large $R_{xx}$ corresponding to the partially filled LLs. We examine the temperature dependence of a specific $(\nu_{LL}^{(h)}, \nu_{LL}^{(e)}) = (1.5, -1.5)$ insulating state at different E-fields in Fig. 5b and Extended Data Fig. 6. The insulating state quickly strengthens under a small E-field and then weakens gradually with increasing E-field (Fig. 5a and 5c). A similar trend is also observed for the other insulating states at half fillings (e.g. $\nu_{LL}^{(h)} = -\nu_{LL}^{(e)} = 2.5$).

The integer-filled states at $\nu_{LL}^{(h)} = -\nu_{LL}^{(e)} = 1,2,3 ...$ are the QSHIs illustrated in Fig. 1c and 1d. Each pair of fully filled electron- and hole-LLs contributes a pair of spin-polarized (due to spin-valley locking), counterpropagating helical edge states; each helical edge state pair contributes a conductance $\frac{e^2}{h}$. Therefore, a standard Hall bar geometry measurement (as employed in our study) gives $R_{xx} = \frac{h}{2\nu_{LL}^{(h)} e^2}$ and $R_{xy} = 0$. (Note that unlike quantum Hall states, the quantization in QSHIs is typically less robust because helical edge states are much more susceptible to back scattering[45,60-63] compared to chiral edge states.) Compared to time-reversal symmetry protected topological insulators, in which only a single pair of helical edge states is allowed, the QSHIs here are protected by spin-$S_z$ conservation and are robust against a perpendicular B-field; multiple pairs of helical edge states are therefore allowed[46,47]. To further support the emergence of QSHIs, we carry out nonlocal transport measurements[64] in Fig. 4d. We fix the B-field at 9.6T and scan the E-field across the transition from the $(\nu_{LL}^{(h)}, \nu_{LL}^{(e)}) = (1, -1)$ QSHI to the $(\nu_{LL}^{(h)}, \nu_{LL}^{(e)}) = (2, -2)$ double QSHI. The measurement configurations are labeled by the same color as the nonlocal resistance ($R_{NL}$) curves. The dashed lines mark the expected quantized values for $R_{NL}$ based on a Landauer-

Büttiker analysis[65] on the nonlocal geometry (Methods). A reasonable agreement between experiment and theory is observed.

Next, we discuss the half-filled states. Without Coulomb interactions, the half-filled states would be a metallic electron-hole plasma. The emergent insulating states are inconsistent with this single-particle picture, but consistent with the intervalley coherent states driven by electron-hole interactions[39]. These intervalley coherent states can also be viewed as EIs as electrons and holes are bound to form excitons spontaneously, resulting in a charge insulating state but an exciton superfluid[11] at low temperatures. Our experiment is, however, unable to establish superfluidity; future studies, such as examination of the proposed spin superfluidity[39], are required to demonstrate the emergence of spontaneous phase coherence. The stability of the EIs is sensitive to Fermi surface nesting effects, which can explain the observed E-field dependence in Fig. 5. In the relevant E-field range, the calculated electron Fermi surface is always a single $\gamma$-pocket (in the K-valley) although the trigonal warping effect becomes more significant at high E-fields (Fig. 5d). On the other hand, the hole Fermi surface consists of two $\kappa_+$- and $\kappa_-$-pockets (in the K'-valley) near zero E-field; a sufficiently high E-field polarizes all the holes to either the $\kappa_+$- or $\kappa_-$-pocket (depending on the sign of the E-field), whose shape becomes weakly dependent on E-field. The evolution of the electron and hole Fermi surfaces with E-field (Fig. 5d) shows that the best Fermi surface nesting condition is achieved at E-fields right after full polarization to $\kappa_+/\kappa_-$. This picture explains the absence of EIs near zero E-fields (because of the highly mismatched electron and hole Fermi surfaces) and the weakened EI state at high E-fields (due to less nested Fermi surfaces).

Finally, we note that with decreasing B-field (below about 7T) the QSHIs give way to the EIs as shown in Fig. 3c. This can be explained by the competition between the cyclotron energy and the exciton binding energy[48,49]. In particular, a higher cyclotron energy relative to the exciton binding favors the QSHIs and vice versa; with sufficiently strong exciton binding (relative to the cyclotron energy), EIs are stabilized even for integer-filled LLs. The behavior is similar to the recently reported QOs between EIs and layer-decoupled quantum Hall states in electron-hole double layers[52,53], where the cyclotron energy and exciton binding also compete. This picture explains the dominance of EIs (QSHIs) at low (high) B-fields.

**Conclusion and outlook**
In conclusion, we have demonstrated QOs between QSHIs and EIs at half-band-filling in tWSe$_2$. This is achieved by utilizing the large valley g-factor and the narrow moiré bandwidth in the material, which allows the creation of a valley-polarized vacuum state and the tuning of the electron-hole pair density (and chemical potential) by an external B-field. We have also demonstrated the tuning of the Fermi surface nesting condition by an external E-field and its effect on the EI stability. Moreover, we have provided a comprehensive understanding of the fermiology in tWSe$_2$. Many open questions remain. For instance, while mean-field calculations have shown that the periodic QSHI-to-EI transitions are first-order[17,18,48,49] in nature, we have not observed any obvious hysteresis in the scanning parameters; future studies (e.g. compressibility measurements) may help discern the order of the transitions. The results demonstrated in this study also provide a path to realize the proposed fractional QSHIs[66] based on conjugate electron- and hole-LLs in the fractional quantum Hall regime.

## Methods

### Device fabrication

A comprehensive description of the device structure and fabrication process can be found in our earlier study[32]; we only provide a short overview here. The complete device stack was assembled in a single, continuous process using standard dry-transfer techniques[67]. Individual flakes were sequentially picked up from top to bottom using a polycarbonate (PC) film on a polydimethylsiloxane (PDMS) stamp; these include a topmost hexagonal boron nitride (hBN) layer, a narrow strip (1-2μm wide) of top-gate graphite, a thin (~3nm thick) top hBN barrier, one half of a WSe$_2$ monolayer, the other half of WSe$_2$ at the targeted twist angle, a bottom hBN barrier (5-10nm), and a wider strip (~5μm wide) of bottom-gate graphite. The assembled stack was then released onto a Si/SiO$_2$ substrate with prepatterned Ti/Pt (5nm/30nm) electrodes. Additional contact and split gates (5nm/40nm Ti/Pd) were then deposited atop the stack using standard e-beam lithography and e-beam evaporation. A Hall-bar geometry is defined by the combined action of the four gates: the channel region is determined by the overlapping area of the top and bottom graphite gates, while the effective contact electrodes are defined by the contact and split gates. The realization of a high-quality twisted homobilayer device with low moiré disorder critically depends on a small channel area, enabled by using a narrow top-gate graphite.

### Electrical measurements

Electrical transport measurements were performed in a dilution refrigerator (Bluefors LD250), equipped with a 12T superconducting magnet. Siver-epoxy filters (Basel Precision Instruments MFT25) integrated with 2-pole RC filters were mounted on the mixing chamber plate to ensure efficient electron thermalization and strong microwave attenuation. Bias voltages of opposite sign were applied to the contact and split gates to activate the contacts and suppress unwanted parallel conduction channels. A small bias current (<5nA) was maintained to prevent sample heating. Standard low-frequency (5.777Hz) lock-in techniques were employed to measure the four-terminal resistance. The voltage drop at the probe electrodes and the source-drain current were simultaneously recorded. All data were acquired at the base temperature (about 40mK) unless otherwise specified.

### Band structure calculation

To capture the low-energy physics of small-angle tWSe$_2$, we calculated the electronic band structure using a continuum model[34,35]. Based on an effective mass description, this model begins with a parabolic valence band $H = -\frac{\hbar^2 k^2}{2m^*}$, where $\hbar$ is the reduced Planck constant, $m^* \approx 0.45 m_0$ (Ref.[68]) is the effective mass, and $\boldsymbol{k}$ is the wave factor ($m_0$ is the free electron mass). The relevant valence bands are located at the K and K' valleys and carry opposite spins due to the strong Ising SOC. The two valleys can be treated as independent since the moiré lattice constant $a_M \approx 52.07$Å is much larger than the atomic lattice constant $a \approx 3.317$Å (Ref.[69]).

We focus on the K-valley with spin-up, for which the effective Hamiltonian is given by

$$H_\uparrow = U^{-1} \begin{pmatrix} -\frac{\hbar^2 \boldsymbol{k}^2}{2m^*} + \Delta_t(\boldsymbol{r}) & \Delta_T(\boldsymbol{r}) \\ \Delta_T^\dagger(\boldsymbol{r}) & -\frac{\hbar^2 \boldsymbol{k}^2}{2m^*} + \Delta_b(\boldsymbol{r}) \end{pmatrix} U$$

under a passive translational transformation

$$U = \begin{pmatrix} e^{iK^t \cdot r} & \\ & e^{iK^b \cdot r} \end{pmatrix}$$

where $K^t$ and $K^b$ denote the K-point of the top and bottom layer, respectively, and correspond to $\kappa_+$ and $\kappa_-$ in the mBZ.

The formation of a moiré superlattice introduces a spatially modulated pseudomagnetic field $\Delta(r) = (\text{Re}\Delta_T^\dagger, \text{Im}\Delta_T^\dagger, \frac{\Delta_t - \Delta_b}{2})$. In the lowest harmonic approximation, the layer-dependent moiré potential $\Delta_{t,b}$ and the interlayer tunneling $\Delta_T$ can be written as

$$\Delta_{t,b}(r) = \pm \frac{V_z}{2} + 2V \sum_{i=1,3,5} \cos(g_i \cdot r \pm \psi)$$

$$\Delta_T(r) = w \sum_{i=1,2,3} e^{-iq_i \cdot r}$$

where $g_i$ and $q_i$ represent the momentum difference between the nearest plane wave bases within the same layer and across different layers, respectively. We took the parameters $(V, \psi, w) = (9.0 meV, 128°, 18 meV)$ obtained from large-scale density function theory calculations at a twist angle $\theta = 5.08°$ (Ref. [35]). The sublattice potential difference $V_z$ can be written as $V_z = \frac{\varepsilon_{hBN}}{\varepsilon_{TMD}} etE$, where the dipole moment $\frac{\varepsilon_{hBN}}{\varepsilon_{TMD}} et \approx 0.26 e \cdot nm$ is independently determined from the anti-crossing features in the layer-hybridized moire excitons[70,71]. Here, $\varepsilon_{hBN} \approx 3$ and $\varepsilon_{TMD} \approx 8$ are the out-of-plane dielectric constants of hBN and TMD, respectively, and $t \approx 0.7\ nm$ is the interlayer separation.

The effective Hamiltonian for the K'-valley (spin-down) is the time-reversal counterpart of $H_\uparrow$. The resulting band structure of tWSe$_2$ was computed by diagonalizing the Hamiltonian for each valley. Based on this band structure, we further calculated the DOS and Fermi surface topology under zero B-field.

**Landauer-Büttiker analysis**
We employed a simplified Landauer-Büttiker (LB) analysis[65] to calculate the chemical potential distribution across different electrodes in the helical edge state transport regime. The LB formula for a multi-terminal device is given by

$$I_p = \sum_q \frac{2e}{h}(\bar{T}_{p \to q}\mu_p - \bar{T}_{q \to p}\mu_q)$$

where $I_p$ is the current injected into electrode $p$, $\bar{T}_{p \to q}$ is the product of the number of conducting channels $M_{p \to q}$ and the transmission probability $T_{p \to q}$ from electrode $p$ to $q$, and $\mu_p$ is the chemical potential of electrode $p$. The prefactor 2 accounts for spin degeneracy. This formula yields a net current in electrode $p$, with each conducting channel contributing $\frac{2e}{h}(\bar{T}_{p \to q}\mu_p - \bar{T}_{q \to p}\mu_q)$.

In the ($n$,-$n$) QSHI state, there are $n$ pairs of counterpropagating edge channels carrying opposite spins. These edge states exist only between adjacent electrodes, with $n$ channels of one spin state propagating downstream and $n$ channels of the opposite spin state propagating upstream. Due to topological protection, each channel supports perfect transmission ($T = 1$), leading to a current $I_{p \to p+1} = \frac{ne}{h}(\mu_p - \mu_{p+1})$ between adjacent electrodes. The result is equivalent to inserting a resistor with resistance $\frac{h}{ne^2}$ between adjacent electrodes. Despite the non-dissipative nature of the helical edge states, a quantized resistance is developed due to the phase decoherence at the contacts, where counterpropagating edge channels are forced to equilibrate at the same potential. As a result, any measurement configuration in QSHI systems can be simplified into an equivalent resistor network.

In Fig. 4d, three different configurations are used to demonstrate the reliability of the nonlocal measurements and the existence of helical edge states. In the standard Hall-bar geometry (black), two edges connect the source and drain, with an equal number of electrodes on both sides of the channel. This symmetry leads to an equal current flow along both edges, $I_{up} = I_{down} = I/2$. Within the equivalent circuit model, the voltage drop between adjacent electrodes is expected to be $\frac{h}{e^2}\frac{I}{2}$ for a QSHI state and $\frac{h}{2e^2}\frac{I}{2}$ for a double QSHI state, yielding quantized resistances of $\frac{h}{2e^2}$ and $\frac{h}{4e^2}$, respectively. In the other measurement geometries (green, purple and pink), the current is applied between two adjacent electrodes, creating a short edge (with no intervening electrodes) and a long edge (intersected by eight electrodes). This asymmetry results in an unequal current flow with $I_{short} = 9I_{long} = 9I/10$. The nonlocal resistance $R_{NL}$ is $\frac{h}{10e^2}$ for a QSHI state and $\frac{h}{20e^2}$ for a double QSHI state, with the sign of $R_{NL}$ determined by the relative orientation between the edge current flow direction and the polarity of the voltage probe. In our experiment, the measured $R_{NL}$ shows reasonable agreement with these expected values (dashed lines in Fig. 4d), supporting the presence of helical edge state transport.

**Estimation of magnetic susceptibility**

The magnetic susceptibility $\chi \propto \frac{h}{e}\frac{\partial(n_+ - n_-)}{\partial B}$ can be expressed in terms of the carrier density imbalance between the two valleys in TMD materials[55]. Here $n_{\pm}$ denotes the carrier density in the K/K' valley. We provide a lower bound for the magnetic susceptibility by using the saturation B-field corresponding to full valley polarization (orange dashed line in Fig. 3b) via $\frac{h}{e}\frac{\partial(n_+ - n_-)}{\partial B} \approx \frac{h}{e}\frac{n}{B_S} = \frac{E_z}{E_c} = g^* m^*/m_0$. Here $n$ is the carrier density in the fully valley-polarized state, which is proportional to the Zeeman energy $E_z$; $B_S$ is the saturation B-field, which is proportional to the cyclotron energy $E_c$; $g^*$ is the effective g-factor. In the layer-polarized region, $\frac{E_z}{E_c}$ was estimated by counting the number of non-degenerate LLs at the saturation $B$-field, while in the layer-hybridized region we directly calculated $\frac{h}{e}\frac{n}{B_S}$ to obtain the susceptibility. The extracted susceptibility $g^*\frac{m^*}{m_0}$ as a function of $\nu$ and $E$ is shown in Extended Data Fig. 3.

**Acknowledgement**
We thank Liang Fu, Chaoming Jian and Debanjan Chowdhury for helpful discussions.

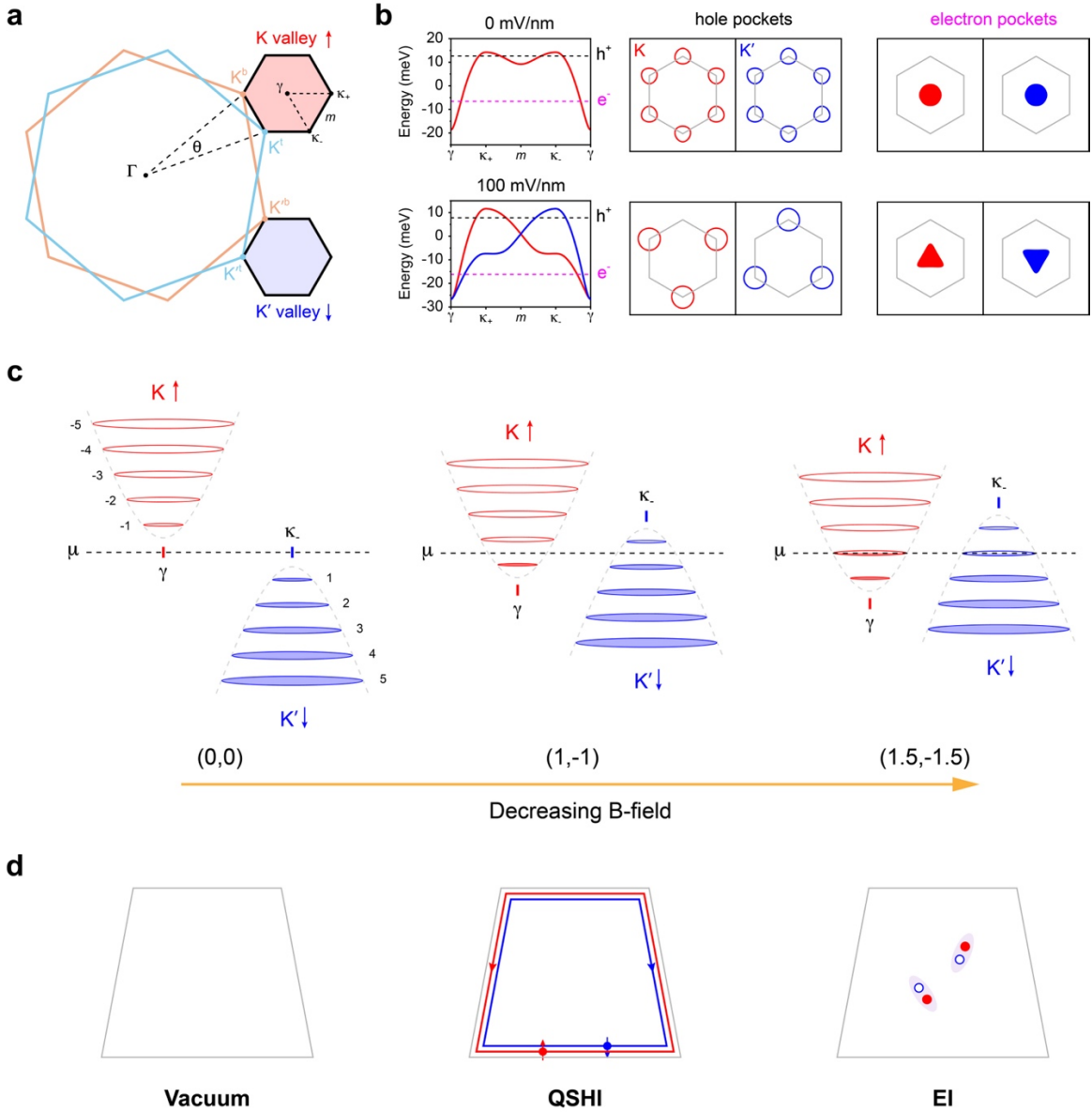

**Figure 1 | Conjugate electron- and hole-Landau levels. a,** Schematic mBZ resulted from the monolayer Brillouin zones at a twist angle $\theta$. The high-symmetry points are labelled. The valley and spin degrees of freedom are locked by the Ising SOC. **b,** Left: The topmost moiré valence bands of 3.65°-twisted tWSe$_2$. Middle (Right): Hole (Electron) Fermi pockets for the Fermi level marked by the black (pink) dashed lines in the left panel. Red and blue colors denote the K- and K'-valley states, respectively. Top and bottom panels correspond to $E = 0$ mV/nm and 100mV/nm, respectively. **c,** Schematic conjugate electron-LLs (at $\gamma$) and hole-LLs (at $\kappa_-$) from opposite valleys under finite E- and B-fields. The dashed line marks the Fermi level at $\nu = 1$, which corresponds to the charge neutrality point. The LLs are fully valley-split in the high B-field limit (left); the hole- and electron-LLs are empty ($\nu_{LL}^{(h)} = \nu_{LL}^{(e)} = 0$), corresponding to a vacuum state shown in **d**. The LLs are occupied with decreasing B-field. The system is expected to be a QSHI at $(\nu_{LL}^{(h)}, \nu_{LL}^{(e)}) = (1, -1)$ (middle) and an EI at $(\nu_{LL}^{(h)}, \nu_{LL}^{(e)}) = (1.5, -1.5)$ (right).

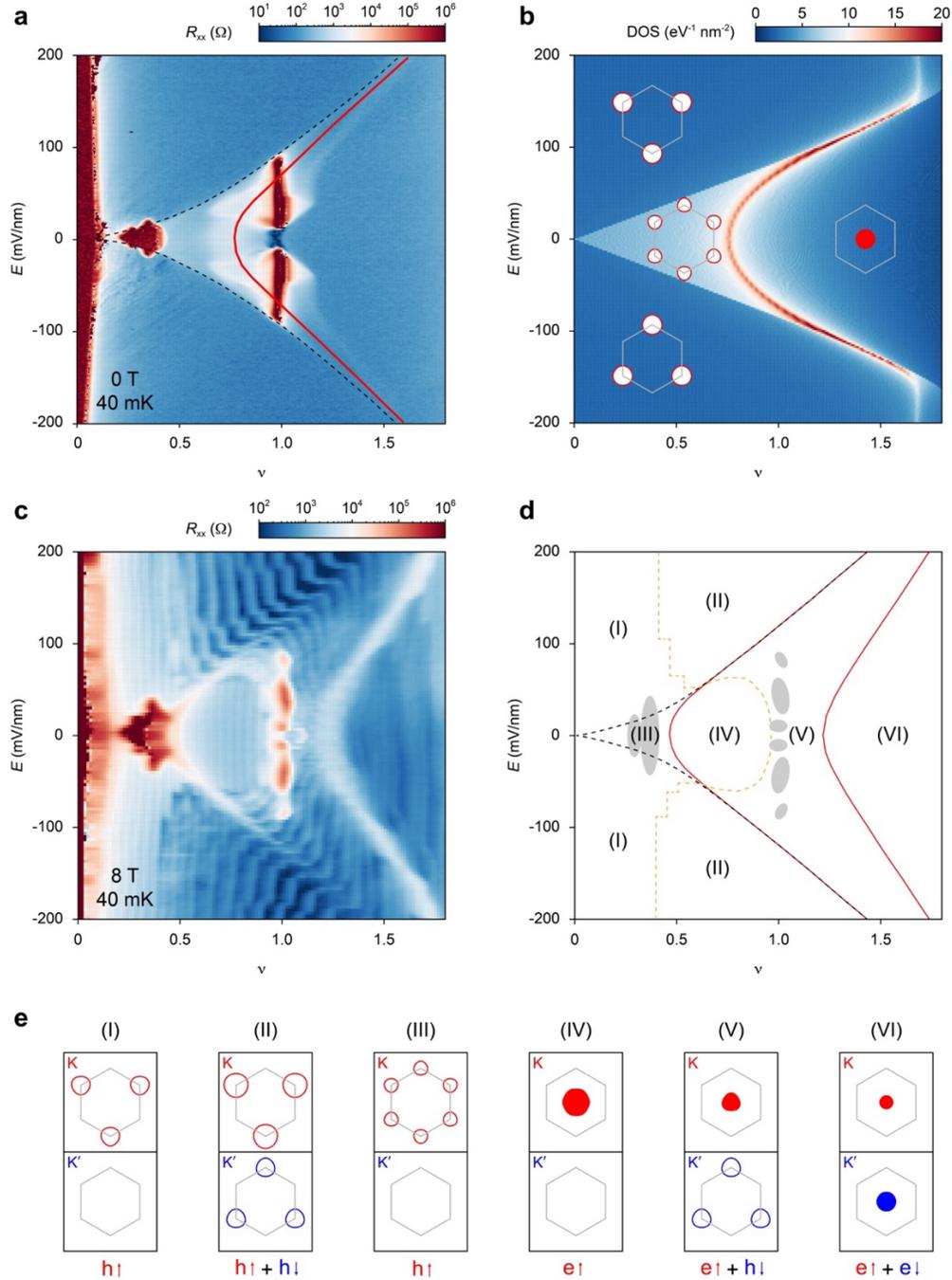

**Figure 2 | Fermiology of tWSe$_2$. a,b,** Longitudinal resistance $R_{xx}$ (**a**) and calculated DOS (**b**) as a function of $\nu$ and $E$ at zero B-field and $T = 40$mK. Representative Fermi pockets (from the K-valley only) are shown for the different regions in **b**. **c,** Same as **a** at $B = 8$T. **d,** Phase diagram constructed from **c**. In **a** and **d**, the black dashed lines separate the layer-polarized and layer-hybridized regions; the red curves trace the location of the vHS, which separates the electron-like and hole-like regions. In **d**, the vHS is valley-split by the external Zeeman field; the orange dashed line separates the fully and partially valley-polarized regions; the grey areas denote the correlated insulating states. **e,** Fermi surface topologies corresponding to the labeled regions in **d**. Empty and filled areas represent hole and electron pockets, respectively.

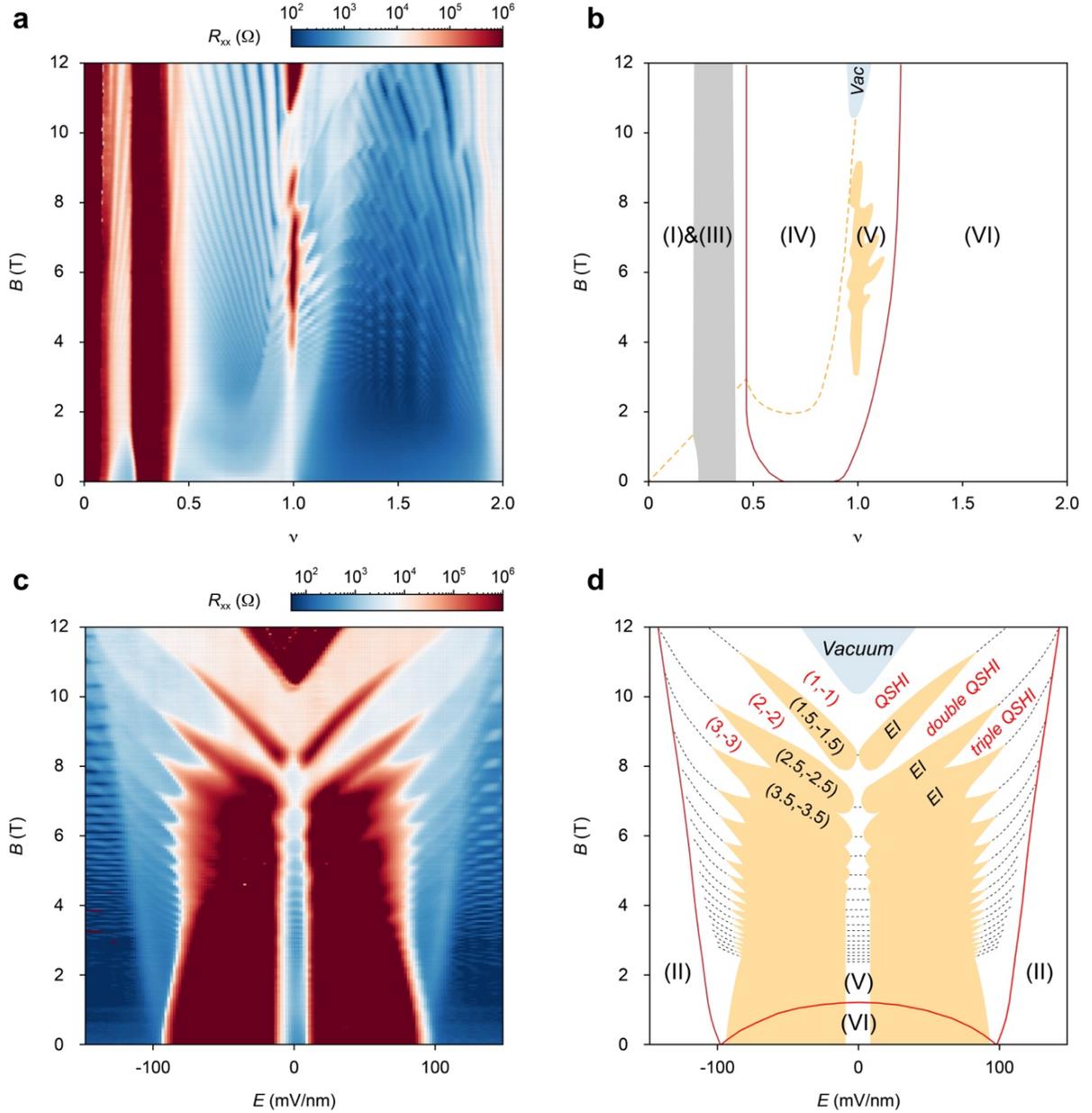

**Figure 3 | Quantum oscillations in correlated insulating states. a,c,** Longitudinal resistance $R_{xx}$ at $T = 40$mK as a function of $\nu$ and $B$ at $E = 10$mV/nm (**a**) and as a function of $E$ and $B$ at $\nu = 1$ (**c**). **b,d,** Phase diagrams corresponding to **a** and **c**, respectively. The vHS (red curves) splits into two branches under B-field; the orange dashed line separates the fully and partially valley-polarized regions; the Fermi surface topologies at different regions are numbered according to **Fig. 2e**. Region V of interest consists of a single K-valley electron pocket centered at $\gamma$ and a single K'-valley hole pocket centered at $\kappa_-$ (under a small E-field). The grey and yellow shaded areas denote the correlated insulating states at $\nu = 1/3$ and $\nu = 1$, respectively; the fully valley-polarized vacuum state at $\nu = 1$ and high B-field is denoted by the blue shaded area. With decreasing B-field, hole- and electron-LLs are sequentially filled while maintaining charge neutrality, i.e. $\nu_{LL}^{(h)} = -\nu_{LL}^{(e)}$. QSHIs and EIs emerge at the fully and partially filled LLs, respectively.

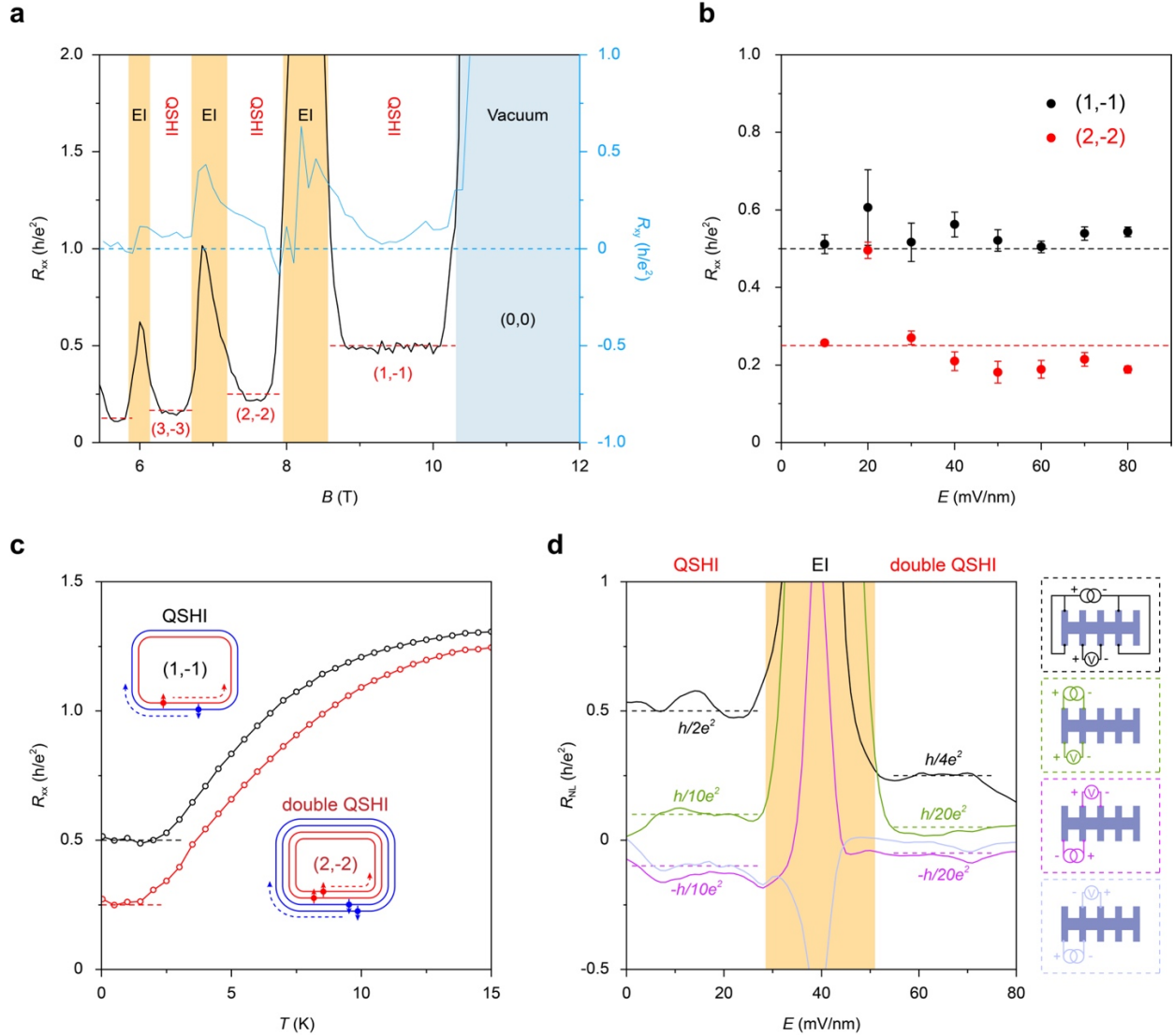

**Figure 4 | QSHIs at fully filled LLs ($\nu = 1$). a,** $R_{xx}$ and $R_{xy}$ as a function of B-field at $E = 5$ mV/nm. EIs and the vacuum states with large $R_{xx}$ and QSHIs with nearly quantized $R_{xx} = \frac{h}{2\nu_{LL}^{(e)}e^2}$ at $\nu_{LL}^{(h)} = -\nu_{LL}^{(e)} = 1,2,3,4$ are labeled. **b,** E-field dependence of $R_{xx}$ for both the single and double QSHIs. Note that each data point corresponds to a value of $R_{xx}$ averaged over a range of B-field where $\nu_{LL}^{(h)} = -\nu_{LL}^{(e)} = 1,2$ (see **Fig. 3c,d**). **c,** Temperature dependence of $R_{xx}$ for the single (black) and double (red) QSHIs. The helical edge states are schematically illustrated in the inset. **d,** Left: E-field dependence of the nonlocal resistance $R_{NL}$ for both the single and double QSHIs at $B = 9.6$ T. The yellow shaded area marks the intervening EI state. Right: The corresponding measurement configurations marked by the same colors as the left panel. + and – denote the polarity of the source-drain and voltage probe pairs. The blue area denotes the device channel. Simultaneous reversal of the polarities of the bias current and voltage probe preserves the sign of $R_{NL}$ for the QSHIs (edge-dominated) but reverses it for the EI (bulk-dominated), as illustrated by the pink and purple curves. The sample temperature is at 40mK for panels **a**, **b** and **d**. The horizontal dashed lines in all panels denote the expected quantized values for $R_{xx}$ or $R_{NL}$.

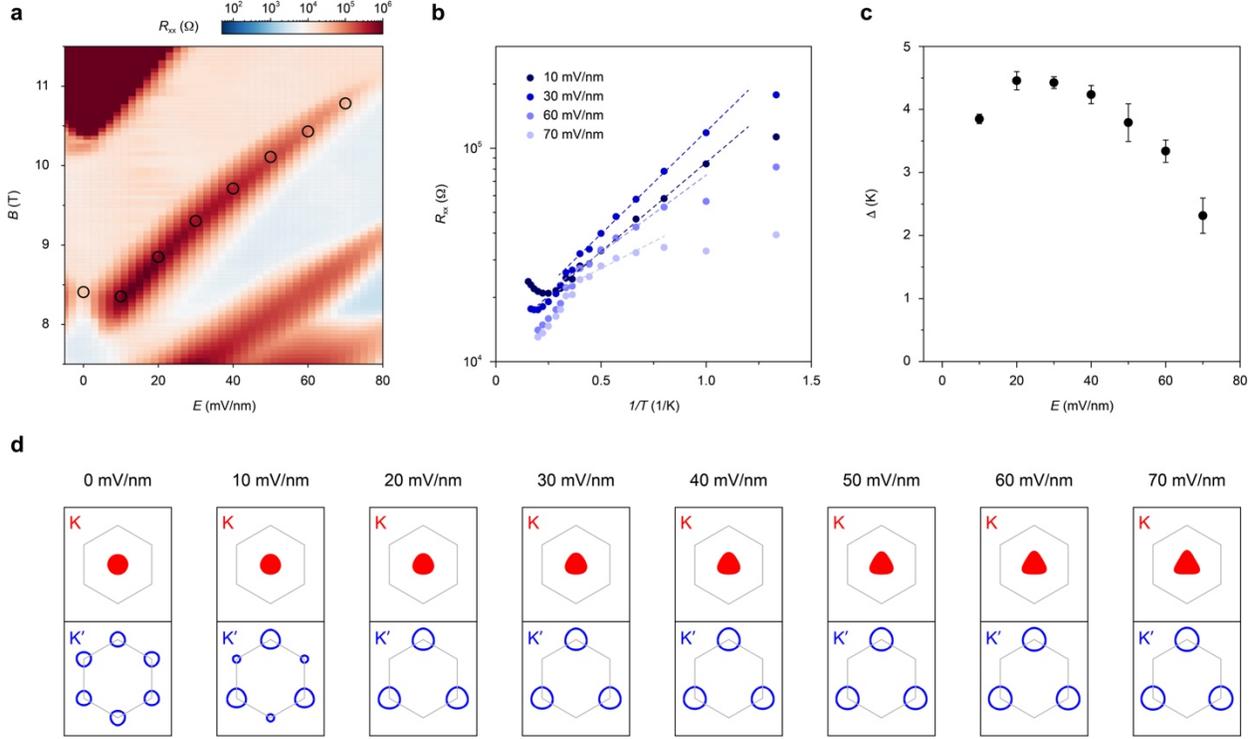

**Figure 5 | EIs at half-filled LLs ($\nu = 1$). a,** Zoom-in $R_{xx}$ map of **Fig. 3c** as a function of $E$ and $B$ at $\nu = 1$, highlighting the $(\nu_{LL}^{(h)}, \nu_{LL}^{(e)}) = (1.5, -1.5)$ EI state. **b,** Arrhenius plot of $R_{xx}$ at the selected locations marked in **a**. Thermal activation fits to the data in the low-temperature regime give the E-field dependence of the EI gap size in **c**. **d,** Fermi surface topologies at different E-fields. Empty and filled areas represent hole and electron pockets, respectively. The Fermi surfaces are calculated for the carrier density $n_{h/e} = \nu_{LL}^{(h/e)} \times \frac{eB}{h}$ at $(\nu_{LL}^{(h)}, \nu_{LL}^{(e)}) = (1.5, -1.5)$.

# Extended Data Figures

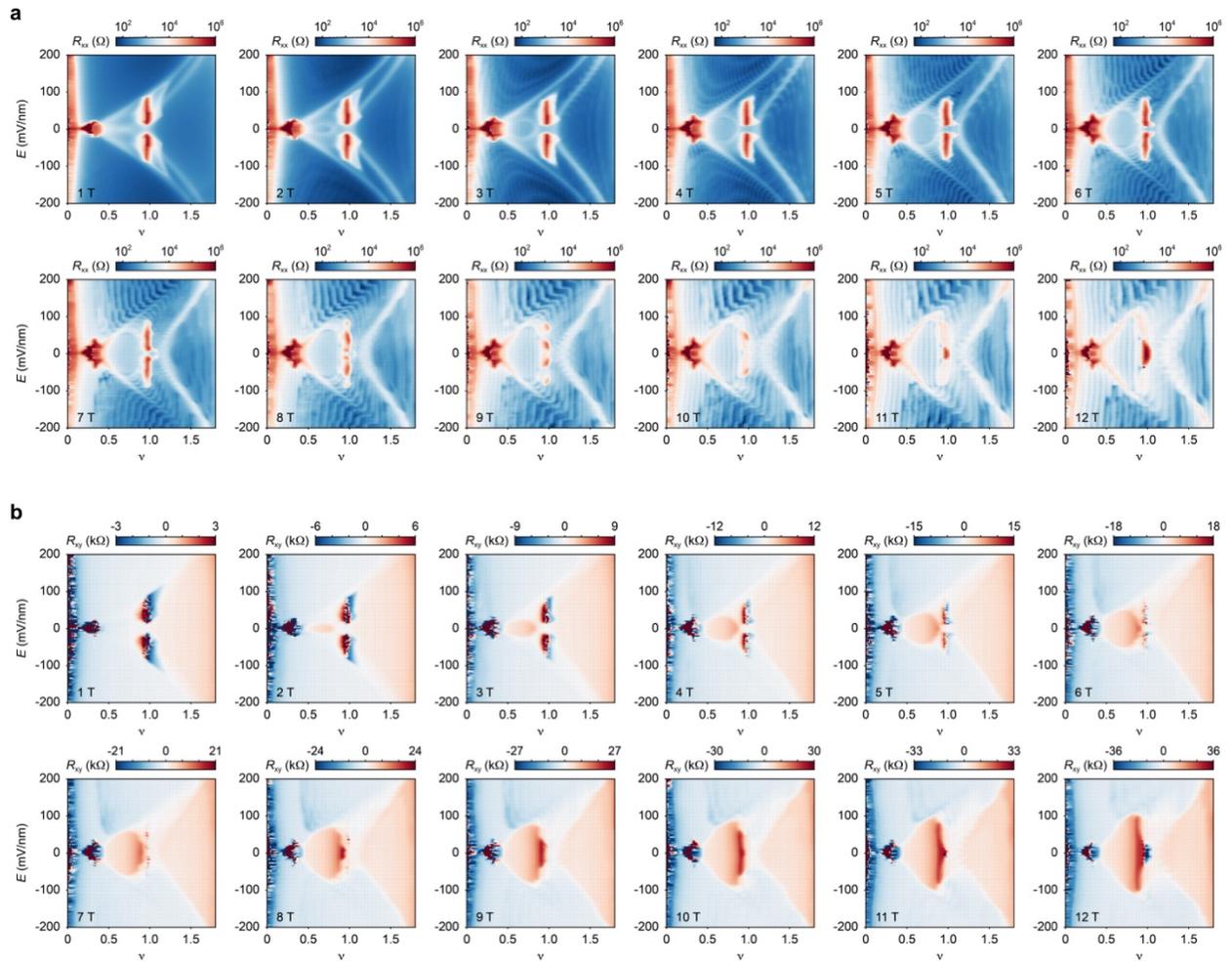

**Extended Data Figure 1 | Additional data at varying B-fields.** Longitudinal resistance $R_{xx}$ (**a**) and Hall resistance $R_{xy}$ (**b**) as a function of $\nu$ and $E$ at varying B-fields ($T = 40$ mK). The phase diagram systematically evolves with B-field. Different regions in the phase diagram can be identified similar to **Fig. 2d**.

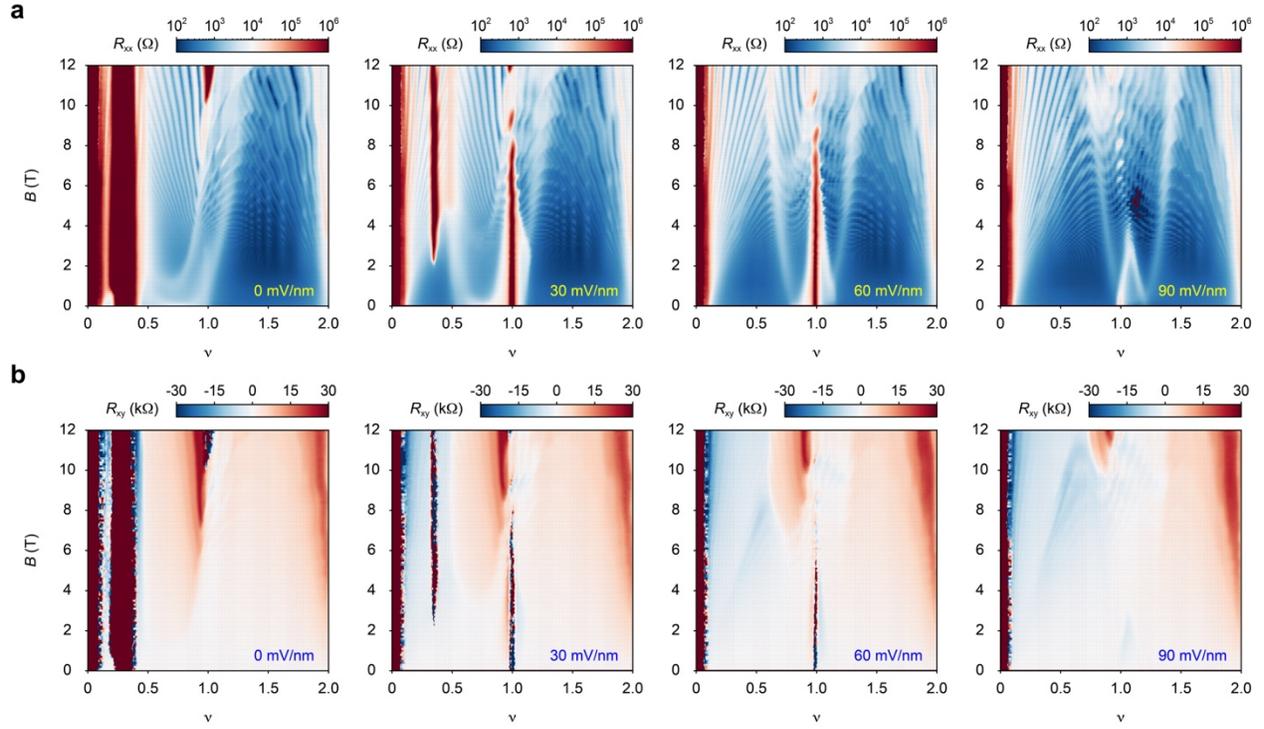

**Extended Data Figure 2 | Additional data at varying E-fields.** Longitudinal resistance $R_{xx}$ (**a**) and Hall resistance $R_{xy}$ (**b**) as a function of $\nu$ and $B$ at varying E-fields ($T = 40$mK). Different regions in the phase diagram can be identified similar to **Fig. 3b**.

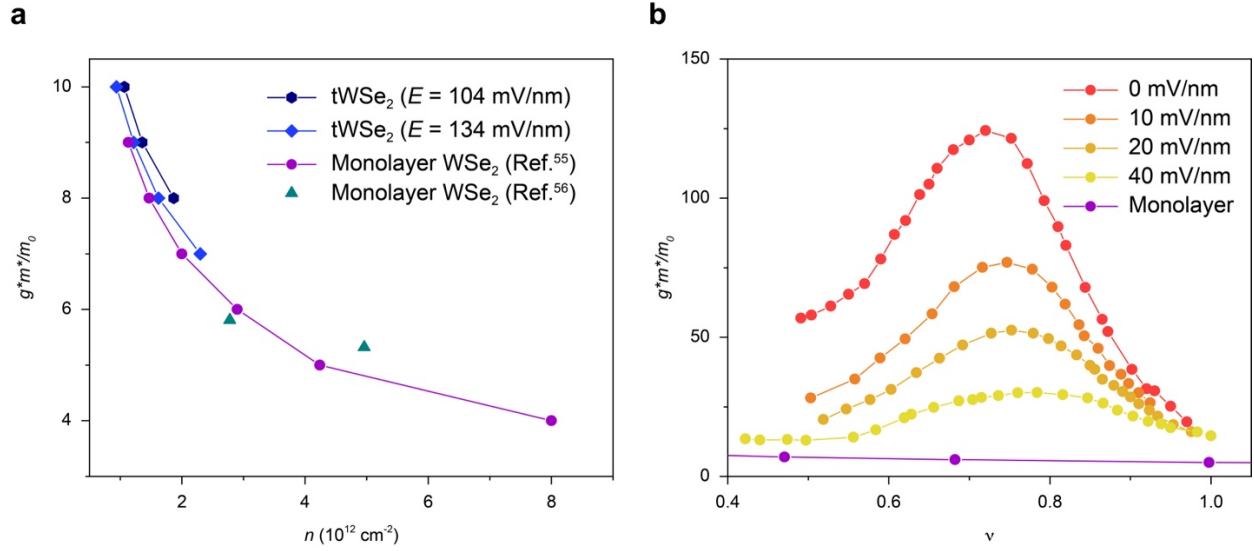

**Extended Data Figure 3 | Doping dependence of the magnetic susceptibility in tWSe$_2$.** Magnetic susceptibility $\chi \propto \frac{E_z}{E_c} = g^*m^*/m_0$ as a function of $\nu$ in the layer-polarized (**a**) and layer-hybridized (**b**) regions ($T = 40$mK). In **a**, where electronic correlations are weak, tWSe$_2$ behaves similarly to monolayer WSe$_2$ (susceptibility data from Ref.[55,56]). The strong electronic correlations in **b** significantly enhance the susceptibility, resulting in a pronounced peak at the vHS near $\nu = 0.75$. As E-field increases, the peak shifts to higher doping densities, tracking the movement of the vHS; the peak enhancement also diminishes.

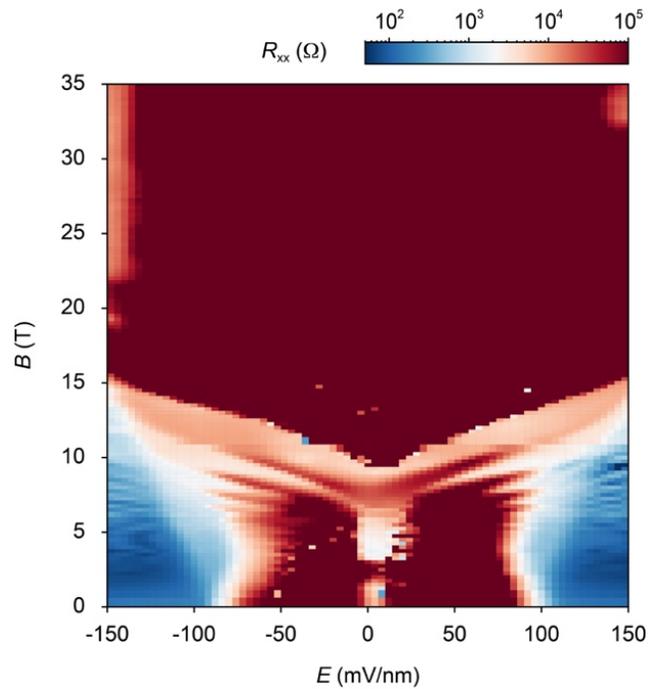

**Extended Data Figure 4 | Magnetoresistance data at high B-fields.** Longitudinal resistance $R_{xx}$ as a function of $E$ and $B$ up to $B = 35$T at $\nu = 1$ ($T = 300$mK). The data is from a second device at a twist angle 3.5°. Whereas the low-field data is similar to that in **Fig. 3c**, the vacuum state dominates the phase diagram for B-fields above $B \approx 10$T.

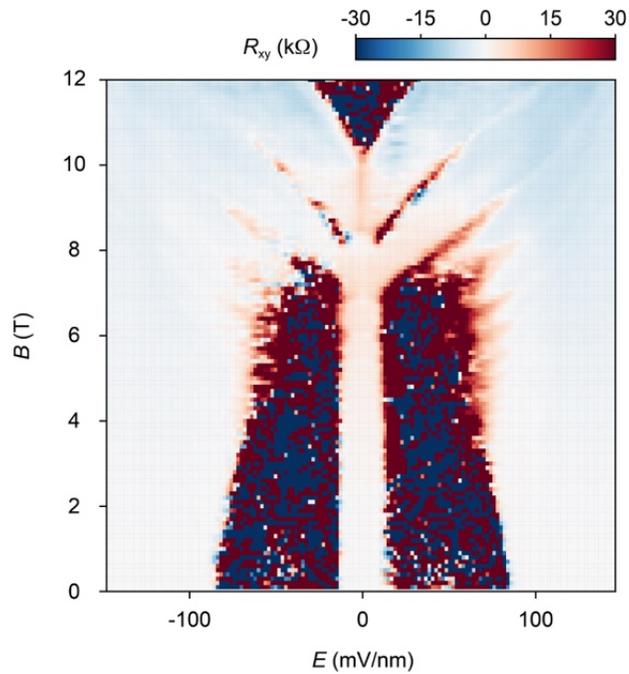

**Extended Data Figure 5 | Hall resistance $R_{xy}$ as a function of $E$ and $B$ at $\nu = 1$ ($T = 40$mK).** The data correlates with the $R_{xx}$ data in **Fig. 3c**. Weak Hall response is observed in the QSHI regions of the phase diagram. Note that reliable Hall data cannot be obtained for the strongly insulating states (e.g. EIs and the vacuum state).

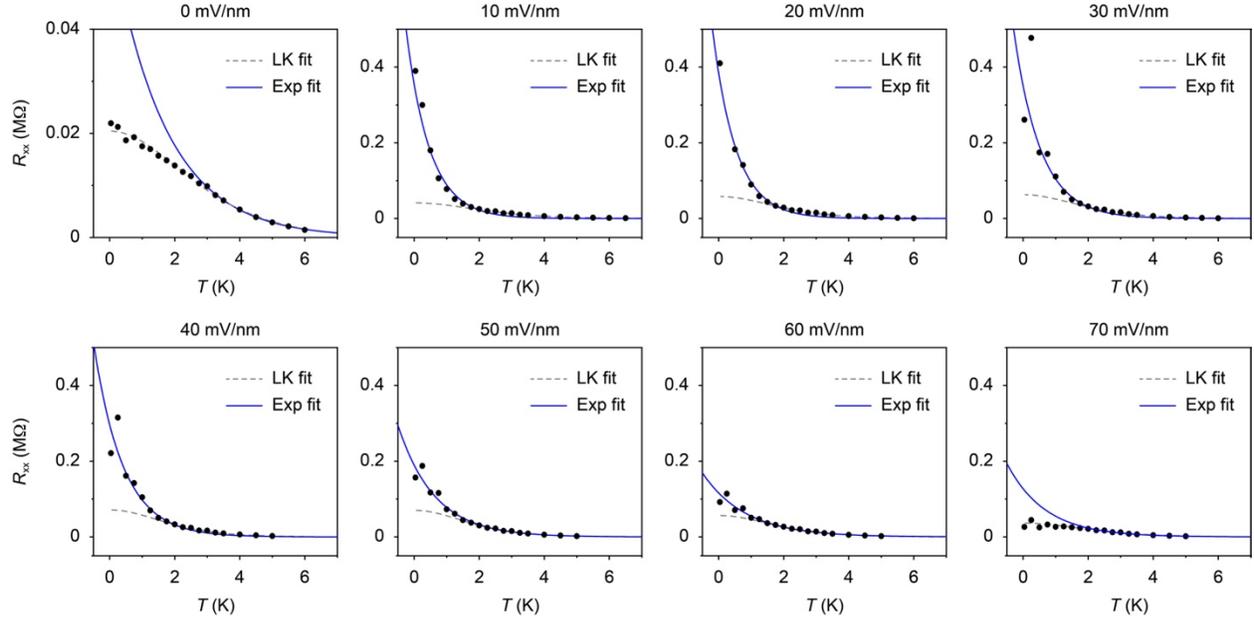

**Extended Data Figure 6 | Temperature dependence of $R_{xx}$ for the EI states at $(v_{LL}^{(h)}, v_{LL}^{(e)}) = (1.5, -1.5)$ and varying E-fields.** Data are obtained at the marked locations in **Fig. 5a**. At zero or sufficiently high E-fields ($E = 70$ mV/nm), where the EI state weakens, the T-dependence follows the Lifshiitz-Kosevich (LK) formula (dashed curves). At intermediate E-fields, where Fermi surface nesting stabilizes the EI state, the T-dependence follows the LK formula at high temperatures but shows strong deviations at low temperatures. In the low-temperature regime, $R_{xx}$ follows an exponential dependence (blue curves), indicating the opening of a correlated insulating gap.